\documentclass[amstex,12pt]{article}
\usepackage{epsf}
\usepackage{amssymb}
\usepackage{cite}
\newlength{\dinwidth}
\newlength{\dinmargin}
\newtheorem{theorem}{Theorem}[section]
\newtheorem{proposition}[theorem]{Proposition}

\def\nind{\noindent}

\def\be{\begin{equation}}
\def\ee{\end{equation}}

\def\RR{{\mathbb R}}
\def\CCC{{\mathbb C}}
\def\IN{{\mathbb N}}

\def\bx{{\mbox{\boldmath{$x$}}}}

\def\bp{{\mbox{\boldmath{$p$}}}}

\def\bq{{\mbox{\boldmath{$q$}}}}

\def\CA{{\cal A}}

\def\CC{{\cal C}}
\def\CH{{\cal H}}

\def\CL{{\cal L}}
\def\CM{{\cal M}}
\def\CN{{\cal N}}
\def\CO{{\cal O}}
\def\CP{{\cal P}}
\def\CR{{\cal R}}

\def\CW{{\cal W}}

\def\AO{{\CA (\CO)}}

\def\cmp{Commun.\ Math.\ Phys.\ }
\def\jmp{J.\ Math.\ Phys.\ }
\def\lmp{Lett.\ Math.\ Phys.\ }
\def\rmp{Rev.\ Math.\ Phys.\ }
\begin{document}
\title{Algebraic Quantum Field Theory: \\ A Status Report}
\author{Detlev Buchholz\thanks{Plenary talk given at 
XIIIth International Congress on Mathematical Physics, London} \\[5mm]
Institut f\"ur Theoretische Physik der Universit\"at G\"ottingen,\\
D-37073 G\"ottingen, Germany}
\date{}

\maketitle

\begin{abstract} \noindent 
Algebraic quantum field theory is an approach to relativistic 
quantum physics, notably the theory of elementary particles, 
which complements other modern developments in this field.
It is particularly powerful for structural analysis but has also
proven to be useful in the rigorous treatment of models. 
In this contribution a non--technical survey is given 
with emphasis on interesting recent developments and future 
perspectives. Topics covered are the relation between the 
algebraic approach and conventional quantum field theory,   
its significance for the resolution of conceptual problems 
(such as the revision of the particle concept)
and its role in the characterization and possibly also 
construction of quantum field theories with the help of 
modular theory. The algebraic approach has also shed 
new light on the treatment of quantum field theories on curved 
spacetime and made contact with recent developments in 
string theory (algebraic holography).
\end{abstract}


\section{Introduction}
In the present year 2000 we are celebrating the 100th birthday of 
quantum theory and the 75th birthday of quantum mechanics. Thus 
it took only 25 years from the first inception of the new theory
until its final consolidation. Quantum field theory is almost as old
as quantum mechanics. But the formulation of a fully
consistent synthesis of the principles of quantum theory and
classical relativisitic field theory has been a long and agonizing
process and, as a matter of fact, has not yet come to a satisfactory end,  
in spite of many successes.

The best approximation to nature in the microscopic and relativistic
regime of elementary particle physics which we presently have, the 
so called {\it Standard Model}, does not yet have the status of a 
mathematically consistent theory. It may be regarded as an efficient 
algorithm for the theoretical treatment of certain specific problems in 
high energy physics, such as the perturbative calculation of collision cross 
sections, the numerical analysis of particle spectra
etc. Yet nobody has been able so far to prove or disprove that the 
model complies with {\it all\/} fundamental principles of relativistic quantum 
physics. 

This somewhat embarassing situation is not so widely known.
It is therefore gratifying that the Clay Mathematics Institute has recently
drawn attention to it by endowing a price of 1.000.000\,\$ for the 
mathematical consolidation of an important piece of the standard 
model, the Yang--Mills--Theory. So the mathematical 
and conceptual problems of relativistic quantum field
theory are in many respects a rewarding field of activity for 
mathematical physicists. 

There are two strategies to make further progress.
Either one tries to improve the existing mathematical methods 
for the treatment of models of physical interest.
This is the approach of constructive quantum field theory
\cite{Ja}. Or one proceeds from a sufficiently rigid   
mathematical framework, consistent with all basic principles 
of quantum field theory, and aims at new conceptual insights 
and constructive ideas by structural analysis.

Such a general framework which is useful for the 
solution of both, conceptual puzzles and constructive problems, 
is {\it Algebraic Quantum Field Theory\/} (AQFT), 
frequently also called {\it Local Quantum Physics}. 
It was invented by Rudolf Haag and
Daniel Kastler \cite{HaKa} and has proven to be consistent with 
the developments in elementary particle physics for several decades.
It is the aim of this contribution to recall the 
physical ideas and mathematical structures underlying AQFT and 
to outline some recent interesting results which reveal its 
flexibility for the treatment of a variety of problems. More 
systematic recent reviews of this approach can be found in \cite{BuHa,Bu}.


\section{Foundations of AQFT}
The relation between the conventional approach to quantum field
theory, based on the Lagrangian formalism, and algebraic 
quantum field theory may be 
compared with the concrete and abstract approaches to 
differential geometry. If one is dealing with concrete (computational)
problems in geometry, it is natural to use coordinates, tensor fields,
Christoffel symbols etc, whereas in the general structural analysis one 
relies on intrinsic concepts such as the notions of manifold, fiber bundle,
connection etc. Both points of view have their virtues and full
insight is only gained by combining them.

The situation is similar in relativistic quantum field theory.
In the concrete Lagrangian approach, one specifies the field 
content of the theory acting on the given space--time manifold $\CM$ 
as well as a corresponding Lagrangian. The difficult step is
``quantization'' which is accomplished either by 
appealing to the correspondence principle (canonical quantization) or 
to general structural results according to which the problem can be  
reformulated in terms of classical statistical field theory 
(Euclidean approach). If everything has been said and done, one
obtains the vacuum correlation functions of the fields. From them
one can reconstruct a Hilbert space on which the observables, 
such as the stress energy tensor, the currents etc.\   
act as operators. The Lagrangian approach is well suited for the computational
treatment of concrete models, but it is not intrinsic. For different 
Lagrangians with different field content may describe the same 
physics. This phenomenon of ``quantum equivalence'' of classical 
field theories has been observed in many examples \cite{Co,Lu,SeWi}.
One may therefore ask whether there is a more intrinsic way of 
describing relativistic quantum field theories.

A fully satisfactory answer to this question is provided by AQFT. 
In this approach the basic objects are the algebras 
generated by the observables localized in given space--time regions;  
fields are not mentioned in this setting and are 
regarded as a kind of coordinates
of the algebras. The passage from the field theoretic setting to the 
algebraic one requires the following steps:
\begin{itemize}
\item[$\triangleright$] determine the set $\{\,\theta (x)\,\}$ of observables 
of the underlying theory for each space--time point $x$ 
\item[$\triangleright$] construct for each relatively compact
space--time region $\CO \subset \CM$ 
a corresponding {\it local algebra of observables}, 
$$ \AO \equiv 
\{\, \theta (x): x \in \CO \,\}^{\prime \prime}, $$
i.e.\ the von Neumann algebra (double commutant)
generated by the respective observables in the 
underlying Hilbert space.
\end{itemize}
In view of the fact that the observables $\theta(x)$ are only defined
in the sense of sesquilinear forms (or as operator valued
distributions), the latter step is somewhat subtle. But it has been
shown to be meaningful in the models which have been constructed so
far \cite{DrFr} and also in the general Wightman setting of quantum field
theory under some very general conditions \cite{BoYn}.

The resulting structure is an assignment of algebras to spacetime regions,
$$\CO \, \mapsto \, \CA(\CO),$$
which is called a {\it local net\/} in view of its order preserving properties.
Any such net inherits some fundamental properties from the underlying field
theory. These are \cite{Ha} 
\begin{itemize}
\item[$\triangleright$] locality: operators localized in causally
  disjoint regions commute
\item[$\triangleright$] covariance: the space--time symmetries act by
  automorphisms on the net 
\item[$\triangleright$] stability: there exist distinguished states
  (expectation functionals) on the net, describing stable elementary
  systems such as the vacuum.
\end{itemize}
Whereas the first two points are well understood and have an obvious 
physical interpretation, the mathematical characterization of
elementary states on arbitrary space--time manifolds is a more difficult
issue which is still under discussion, cf.\ \cite{Wa,BrFrKo,BuDrFlSu,Fe,SaVe} 
and references quoted there. For the class of maximally symmetric
spaces there are no such problems, however.

According to the deep insights of Haag and Kastler, the
full physical information of a theory is already contained in the net
structure, i.e.\ the respective map from space--time regions to 
algebras. Phrased differently, equivalent quantum field theories can 
be identified by the fact that they generate isomorphic local nets.
This assertion may be somewhat surprising at first sight since the 
passage from the observables, which normally have a specific physical
interpretation, to the local algebras seems to be a rather forgetful
operation. That no information is lost in this step has been
confirmed by now by numerous results \cite{Ha}. 

A direct way of seeing this has been established by Fredenhagen and
Hertel \cite{FrHe} who proved for the class of 
Minkowski space theories that the 
set of basic observables can be recovered from the local net by the formula
$$\{\phi(x)\} = \bigcap_{\CO \supset x} \overline{\CA(\CO)}.$$
The somewhat tricky point in this reconstruction is the need to 
proceed from the algebras $\CA(\CO)$ of bounded operators to unbounded
sesquilinear forms. It is accomplished by completing these algebras 
in a suitable locally convex topology, indicated by the bar. 
It should be noted, however, that the algebraic framework is in some
respect more flexible than
the field theoretic setting. For the local algebras may also
accommodate extended objects, such as Wilson loops or finite Mandelstam
strings, which are not built from point like observable fields.

AQFT is thus compatible with the structures
found in the quantum field theories of present physical interest
and complements them by putting emphasis on their 
intrinsic features. It may thus be regarded as a minimal
setting for the description of the systems appearing in high energy
physics. We discuss in the following some issues where the virtues of
this approach become manifest.


\section{Perturbative AQFT}
Guided by insights gained from algebraic quantum field theory, 
Brunetti and Fredenhagen \cite{BrFr} have recently established a perturbative
construction of local nets on arbitrary globally hyperbolic
space--times $\CM$. They propose to treat this problem
in two steps: first one constructs the nets of local
algebras in some convenient Hilbert space representation,
thereby fixing the theory. This step requires control of the notorious  
ultraviolet divergences (renormalization) and can be handled by
configuration space methods and microlocal techniques.
Infrared problems, related to the so--called adiabatic or infinite 
volume limit, do not appear in this construction.
In a second step one may then turn to the determination of the 
states of physical interest on this net and to their analysis. 
This requires the passage to new Hilbert space representations of 
the algebraic structures and may thus be regarded as a problem in the 
representation theory of local nets. 

In Minkowski space theories both problems are frequently treated
simultaneously because of the possibility of characterizing the vacuum
state directly by momentum space properties (spectrum condition). But,
as indicated above, this strategy does not work 
for arbitrary space--time manifolds. Thus the Brunetti--Fredenhagen
approach is a very natural way of circumventing these difficulties.

Following \cite{BrFr}, we outline this method by discussing the 
theory of a self-interacting scalar field on a given space-time $\CM$.
The perturbative construction of the net requires the following steps:
\begin{itemize}
\item[$\triangleright$] consider the free scalar field 
$\phi_{\, 0}$ on the space--time $\CM$,  
$$ 
(\square + m^2) \, \phi_{\, 0}(x)=0, \qquad 
[\phi_{\, 0}(x), \phi_{\, 0}(y)] = -i \, \Delta(x,y) \, 1,$$
where $\square$ denotes the D'Alembertian and $\Delta(x,y)$ the 
causal commutator function on $\CM$, 
in a regular Hilbert space representation induced by some 
Hada\-mard state \cite{Wa}. 
\item[$\triangleright$] construct {\it Wick powers\/} 
  of the free field for $n \in \IN$,
$$  : \! \phi_{\, 0}^{\, n} \!:(x).$$
\end{itemize}
The existence of these operator--valued distributions was first
established in \cite{BrFrKo} by methods of microlocal analysis 
and shown in \cite{BrFr} to be largely independent of
the chosen regular Hilbert space representation. These Wick powers are the
building blocks for the construction of interacting fields.
\begin{itemize}
\item[$\triangleright$] define the time ordered exponentials
(local S--operators)
$$ S(g) \equiv T \exp(\, i \! \int \! d\mu(x)
\, g(x) : \! \phi_{\,  0}^{\, n} \! : \! (x) \,),$$
where $g$ is any test function and $d\mu(x)$ is the volume form on
$\CM$. (More generally, one considers such exponentials for  
finite sums of Wick powers.)
\end{itemize}
The proof that these exponentials are meaningful expressions  
is the most difficult step in the construction. It has 
been established in \cite{BrFr} in perturbation theory by
defining $S(g)$ as formal power series in $g$. The  
coefficients in this series suffer from  
ambiguities due to short distance singularities
which require renormalization. This problem is solved by generalizing 
methods of Epstein and Glaser (causal perturbation theory). If 
$n \leq 4$, there does not appear a proliferation of these 
ambiguities with increasing order of perturbation theory (renormalizability). 
\begin{itemize}
\item[$\triangleright$] use Bogolubov's formula
$$ S_g (f) \equiv S(g)^{-1} S(f+g), \ \ g \upharpoonright \mbox{supp} f = q$$ 
to define, for given interaction density 
$q  : \! \phi_{\, 0}^{\, n} \! : \! (x)$, 
local operators for the cutoff density   
$g (x) : \! \phi_{\, 0}^{\, n} \! : \! (x) $.
\end{itemize}
Up to this point the construction is akin to the treatment of
Minkowski space theories, although the technical details are 
more involved in view of the absence of space--time symmetries. 
The adiabatic limit of $S_g (f)$ for $g \rightarrow q $ may not exist,
however, due to infrared problems caused by the {\it ad hoc\/} 
choice of a defining representation of the interacting theory.
This difficulty can be circumvented by the following important
observation \cite{BrFr}.
\begin{proposition} 
Let $\CO \subset \CM$ and let
$g_1 \upharpoonright \CO = g_2 \upharpoonright \CO = q$. There
exists a unitary operator $V_\CO$ such that 
$$ S_{g_2}(f) = V_\CO S_{g_1}(f) V_\CO^{-1}, \quad 
\mbox{supp} f \! \subset \! \CO.$$
\end{proposition}

\vspace*{2mm}
\nind In view of this result it is meaningful to define local algebras,
setting for $\CO \subset \CM$ and any $g$ with $g \upharpoonright \CO = q$ 
$$\CA_g (\CO) \equiv \mbox{*--algebra} \, \{ S_g(f): 
\mbox{supp} f \! \subset \! \CO \}.$$
According to the preceding proposition, these algebras are unique up 
to isomorphisms $\mbox{ad}V_\CO$, which do not change the physical 
interpretation. One may thus proceed to an algebraic adiabatic 
limit by considering the algebras 
$$\CA (\CO) \equiv \mbox{*--algebra} \, \{ \big( S_g(f) 
\big)_{g \upharpoonright \CO = q} :  \mbox{supp} f \! \subset \! \CO \}.$$
The inclusion (net) structure of this family of algebras 
is given by their natural embeddings and
the algebraic operations of addition, multiplication as well as the  
*--operation in $\CA (\CO)$ are pointwise defined for each $g$. 
In this way one arrives at a (perturbative) net
$$
\CO \, \mapsto \, \CA(\CO) 
$$
for the given interaction. 

Thus the algebraic point of view leads to a natural perturbative
construction of 
nets of local algebras in any space--time $\CM$.
Similar methods have also been applied to the construction of local
nets of observables in gauge theories in Minkowski space \cite{DuFr}, cf.\ also
\cite{Sch} and references quoted there.


\section{Particle analysis} 
We turn next to a conceptual problem in Minkowski space quantum field
theories, namely the asymptotic particle interpretation. In the gauge
theories of physical interest, the basic fields are in general
unphysical and not related to stable particles. One has therefore to 
develop methods to determine the particle features of these theories 
from the vacuum correlation functions of the local observables. The 
confinement problem in quantum chromodynamics and the infraparticle
problem  in quantum electrodynamics are two well known examples
illustrating this issue. Here the algebraic point of view led
recently to some interesting  progress. 

\vspace*{2mm}
\nind The following steps are necessary in the particle 
analysis of any theory:
\begin{itemize}
\item[$\triangleright$] introduce some meaningful particle concept
\item[$\triangleright$] establish methods to determine the 
  particle content of the theory
\item[$\triangleright$] analyze the properties of these particles 
\item[$\triangleright$] develop a scattering (collision) theory. 
\end{itemize}

One has a quite satisfactory understanding of these points in
theories with short range forces (mass gap). There the natural starting point
is Wigner's particle concept, according to which the possible states of a 
particle are described by vectors in some irreducible representation
of the Poincar\'e group or its two--fold covering, respectively. It 
is well known, however, that this approach does not work, 
for example, in theories with 
electrically charged particles, cf.\ for example \cite{Bu2}. So 
it seems desirable to develop a universal particle concept which applies
also in those cases.  

Such a more flexible particle concept was introduced in \cite{BuPoSt,Bu3}. It
is based on the notion of {\it particle weight\/} which is a
generalization of Dirac's idea of an improper momentum eigenstate 
of a particle. There are two possibilities of looking at these 
improper states. 
\begin{itemize} 
\item[$\triangleright$] traditional: improper momentum eigenstates are 
  regarded as vector valued
  distributions, i.e.\ maps from a space of wave functions into the
  physical Hilbert space, 
$$ | \ \ \rangle : \ f \mapsto \int \! dp \, f(p) | \, p \,
\rangle \in \CH.$$
It is anticipated in this approach that the improper states become 
normalizable by superposition (interference effects).
\item[$\triangleright$] alternative: improper states of {\it fixed\/} 
  momentum are regarded as linear maps from a space $\CL$ of localizing
  operators into the physical Hilbert space, 
$$ | \, p \, \rangle : \ L \mapsto L | \, p \, \rangle \in \CH.$$
Here the improper states become normalizable by localization. 
\end{itemize}
It is important to notice that the second approach is more general
than the first one. It is expected to be applicable 
even if the superposition principle fails for the improper states.
This happens, for example, if the process of localization 
is inevitably accompanied by particle production  
(such as in quantum electrodynamics, where infinite clouds of
soft photons are produced).  

In quantum mechanics, suitable localizing operators would be rapidly
decreasing functions of the position operator; but the notion of position 
operator is not meaningful in a field theoretic setting.   
Nevertheless, localizing operators $L$ exist in AQFT in abundance 
and are easily constructed. Simple examples are all operators of the form
$$ L = \int \! dx \, f(x) \, A(x).$$
Here $f$ is any test function whose Fourier transform vanishes
in the forward light cone and $A(x) = U(x) A U(x)^{-1}$, where 
$U(x)$ are the unitaries inducing the  
space--time translations $x$ and $A$ is any local observable.  
The operators $L$ can be shown to annihilate all states 
(disregarding vectors of arbitrarily small norm)    
which do not describe excitations of the vacuum in some 
(sufficiently large but finite) space--time region $\CO_L$. 
So, roughly speaking, 
they ``project'' onto states which differ from the vacuum 
in $\CO_L$. In this sense they are localizing operators. It is
technically important that these localizing operators form a left 
ideal $\CL \subset \CA$ in the C$^*$--algebra $\CA$ generated by 
all local observables.

It is convenient to proceed from the improper states to corresponding
positive, linear and non--normalizable functionals on the domain
$\CL^* \CL \subset \CA$, 
$$\langle p | \, \cdot \, | p \rangle \ : \ \CL^* \CL \rightarrow \CCC,$$
called particle weights.
These functionals can be characterized in an intrinsic manner.  
In particular they are extremal and invariant under space--time
translations. As $\CL^* \CL$ is a $*$--algebra, one can recover by
the GNS reconstruction theorem the improper particle states from these 
particle weights. 

After having introduced in AQFT a general particle concept, one has  
to develop methods to determine the particle content of a theory 
(described by particle weights). This is accomplished by analyzing the 
timelike asymptotic properties of the physical states in the vacuum
sector $\CA \, \Omega = \{ A \Omega : A \in \CA \}$ 
of the theory. So let $\omega(\,\cdot\,)$ be any expectation functional
induced by vectors in the subspace 
$\CA \, \Omega$ of the physical Hilbert space $\CH$.
One then considers the functionals 
$$ \rho_t (L^*L) \equiv 
\mbox{\large ${1 \over t}$} \int_t^{\,2t} \! dx_0 
\! \int \! d \bx \
\omega( \, (L^* L) (x) \, ), \ \ L \in \CL.$$
These expressions are mathematically meaningful as a 
consequence of locality and the shape of the energy momentum spectrum 
\cite{Bu4}. Moreover, the family of functionals $\{\rho_t\}_{t\in\RR}$ is 
equibounded. So this family has limit points 
$$ \rho_\infty (L^* L) = 
\mbox{``} {\lim_{\,t \rightarrow \infty}} \mbox{''} \, \rho_t (L^*L), 
\ \ L \in \CL.$$
The functionals $\rho_\infty$ are, by their very construction, 
invariant under translations, but highly mixed. It is therefore 
natural to ask whether they can be decomposed into particle weights. 
An affirmative answer to this
question was recently given by Porrmann \cite{Po}. More
specifically, there holds the following statement. 
\begin{proposition} 
$$ \rho_\infty (L^*L) = \int \! d\mu(p,\iota) \, 
\langle p,\iota | L^* L | p,\iota \rangle, \ \ L \in \CL$$
where each $\langle p,\iota | \, \cdot \,  | p,\iota \rangle$ is a
particle weight of momentum $p$ and  
``internal index'' $\iota$ and $ d\mu(p,\iota)$ is a measure depending on the
initial state $\omega$. (The internal index $\iota$ describes the intrinsic
features of a particle, such as its charge quantum numbers and spin.)
\end{proposition}
Thus the preceding asymptotic construction and subsequent decomposition
of functionals provides a general method to determine the stable particle
content of any theory, including the charged particles (which  
appear in the vacuum sector only as pairs of opposite charge, 
but asymptotically give rise to 
particle weights contributing to the mixtures $\rho_\infty$). 
This result is a substantial 
generalization of work of Araki and Haag for massive theories with 
a complete particle interpretation \cite{ArHa}. No {\it a priori\/} 
input about particles is necessary for its derivation; this 
shows that the concept of particle weight is sufficient for the
description of the asymptotic particle features of any theory. 

The next step in the analysis is the determination of the
possible properties of particle weights. To this end 
one first constructs 
for each particle weight the corresponding sector of the physical
Hilbert space, 
$$
\CH_{p,\iota} \equiv \overline{\CL \, |p,\iota\rangle} \subset \CH. 
$$
One can then establish the following general results \cite{BuPo}.
\begin{itemize}
\item[$\triangleright$] mass: the energy--momentum $p$ of the
  underlying improper state $|p,\iota\rangle$ and therefore its 
  mass $m^2 = p^2$ can be sharply defined in the sector $\CH_{p,\iota}$
  by the formula 
  $$U_{p,\iota}(x) \ L \, |p,\iota\rangle = e^{ipx} \  
  L(x) \, |p,\iota\rangle,$$
  where $U_{p,\iota}(x)$ is the unitary representation of the
  translations on $\CH_{p,\iota}$ (which can be shown to exist).
\item[$\triangleright$] spin: if, for given $p$, there is a {\it finite\/}
  multiplet of particle weights 
  \mbox{$\langle p,\iota |\,\cdot\,| p,\iota \rangle$},
  there exists a unitary representation $U_{p,\iota}$ 
  of the (covering group of the)
  stability group $\CR$ of $p$ such that for $R \in \CR$
  $$U_{p,\iota}(R) \ L \, |p,\iota\rangle =  
  \sum_\kappa D_{\iota  \kappa}(R^{-1}) \, L(R) \, |p,\kappa\rangle,$$
  where $D$ are matrix representations of $\CR$ and $L(R) = U(R) L
  U(R)^{-1}$. Thus if $m > 0$, the
  particle weights can have spin $s=0, 
  \mbox{${1 \over 2}$}, 1, \mbox{${3 \over 2}$} \dots$, in
  accordance with the results found by Wigner. However, in contrast to the
  case of Wigner particles, there may appear for particle weights with mass
  $m=0$ representations with arbitrary helicity. 
  \item[$\triangleright$] coherence: let $
  \CH_{p,\iota} = \int \! d\bq \, \CH_{p,\iota} (\bq) $ be the 
  decomposition of $\CH_{p,\iota}$ with respect to the spatial
  momentum and consider the restrictions  
  $H_{p,\iota} \upharpoonright  \CH_{p,\iota}  (\bq)$
  of the generator $H_{p,\iota}$ of the time translations to the
  respective subspaces. For the point spectrum of these restrictions 
  there appear the possibilities 
$$
 \sigma_{\mbox{\small\it point}} \, 
\{ H_{p,\iota} \upharpoonright  \CH_{p,\iota}  (\bq) \} \neq
\left\{ \begin{array}{c@{\ \ \mbox{:} \ \ }l}
    \emptyset & \mbox{for all} \ \bq \in \RR^3 \\
    \emptyset & \mbox{if and only if} \ \bq = \bp.
\end{array} \right.
$$
\end{itemize}
The former case corresponds to the familiar situation of Wigner
particles, where the improper states $|q,\iota\rangle, \, \bq \in \RR^3,$
are affiliated with the same sector and thus can coherently be
superimposed. In the latter case, where 
$\{ H_{p,\iota} \upharpoonright  \CH_{p,\iota}  (\bq) \}$
has purely continuous spectrum for $\bq \neq \bp$, the 
superposition principle fails for particle weights of different momenta.
This situation is expected to prevail in theories with long range
forces. 

In view of the latter result it seems a rewarding experimental
challenge to test the status of the superposition principle for electrically 
charged particle weights. The theory predicts that there ought to be 
a substantial difference in the asymptotic coherence properties 
of electrically neutral and charged particles.

In a final step one has to establish a collision theory for particles
described by weights. In general one may not expect that a scattering
matrix exists for these entities, but it is always possible to define and
compute collision cross sections. A general method to this effect has been
outlined in \cite{BuPoSt}. 

We conclude this section with the remark that, 
in contrast to the case of many particle quantum mechanics, the
completeness of the particle interpretation in quantum
field theory is still an open problem, even in the case of short range
forces. A survey of the state of
the art can be found in \cite{Ia}. 


\section{Algebraic holography and transplantation}  

Triggered by developments in string theory, known under the 
catchwords of holography or AdS/CFT correspondence (cf.\ \cite{Wi} 
and references quoted there), there has recently emerged interest in 
the relation between quantum field theories on different space--time 
manifolds. 
One speaks of {\it holography\/} if there is a correspondence between a
quantum field theory on a space--time $\CM_1$ and a 
theory on its boundary $\CM_2 = \partial \CM_1$. In other
words, given the theory on $\CM_1$, one can unambiguously determine the
theory on $\CM_2$ and {\it vice versa}. A related notion is
{\it transplantation}, where such a correspondence exists between theories on
space--times $\CM_1$, $\CM_2$ of equal dimension.    

In the conventional field theoretic setting a 
satisfactory understanding of these issues seems impossible.
It is clear from the outset that one cannot 
identify point fields on $\CM_1$ and $\CM_2$ in a meaningful way,  
$$ \phi_{\CM_1} (x_1) \nsim \phi_{\CM_2} (x_2).$$
With regard to holography, the best one can do is to proceed from a
field theory on $\CM_1$ by restriction to a corresponding theory 
on $\CM_2$, cf.\ \cite{BeBrMoSch}. But it is in general impossible 
to recover from the latter data the original fields. Similar problems 
hamper also the idea of transplanting fields. 

Here the algebraic point of view provides a
solution: bearing in mind that a theory is fixed by the underlying
net, one realizes that it is sufficient (and frequently possible) to
identify local algebras for certain sufficiently rich families of
regions $\{ \CO_1 \subset \CM_1 \}$, $\{ \CO_2 \subset \CM_2 \}$,  
$$ \CA_{\CM_1}(\CO_1) \sim \CA_{\CM_2}(\CO_2),$$
thereby establishing a rigid link between the two theories in question.
This insight was first used by Rehren in his analysis of the issue of 
holography \cite{Re}. Similar ideas were applied
to the problem of transplantation in \cite{BuMuSu}. We outline in the
following the underlying simple geometrical facts.

\vspace*{2mm}
\nind Holography: \\ 
The simplest example, where the idea of holography can be illustrated, 
is the correspondence between quantum field theories on anti--de Sitter
space (AdS) and on its Minkowskian boundary. In proper coordinates, AdS
can be envisaged as a full cylinder whose tube like boundary is  
conformal Minkowski space. 

\nind As indicated above, one has to identify suitable regions
in the given space--times in order to establish a correspondence
between the respective theories. For the case at hand, Rehren proposed 
to consider a family of causally complete wedge--shaped regions  
$ \{ \CW \subset \mbox{AdS} \}$. Their intersections with the boundary
of AdS are diamond shaped regions, 
$\{\CC \equiv {\CW} 
\upharpoonright \partial \mbox{AdS} \}$, cf.\ Figure 1. 

\hspace*{6mm}
\begin{figure}[!h]
\epsfxsize140mm
\epsfbox{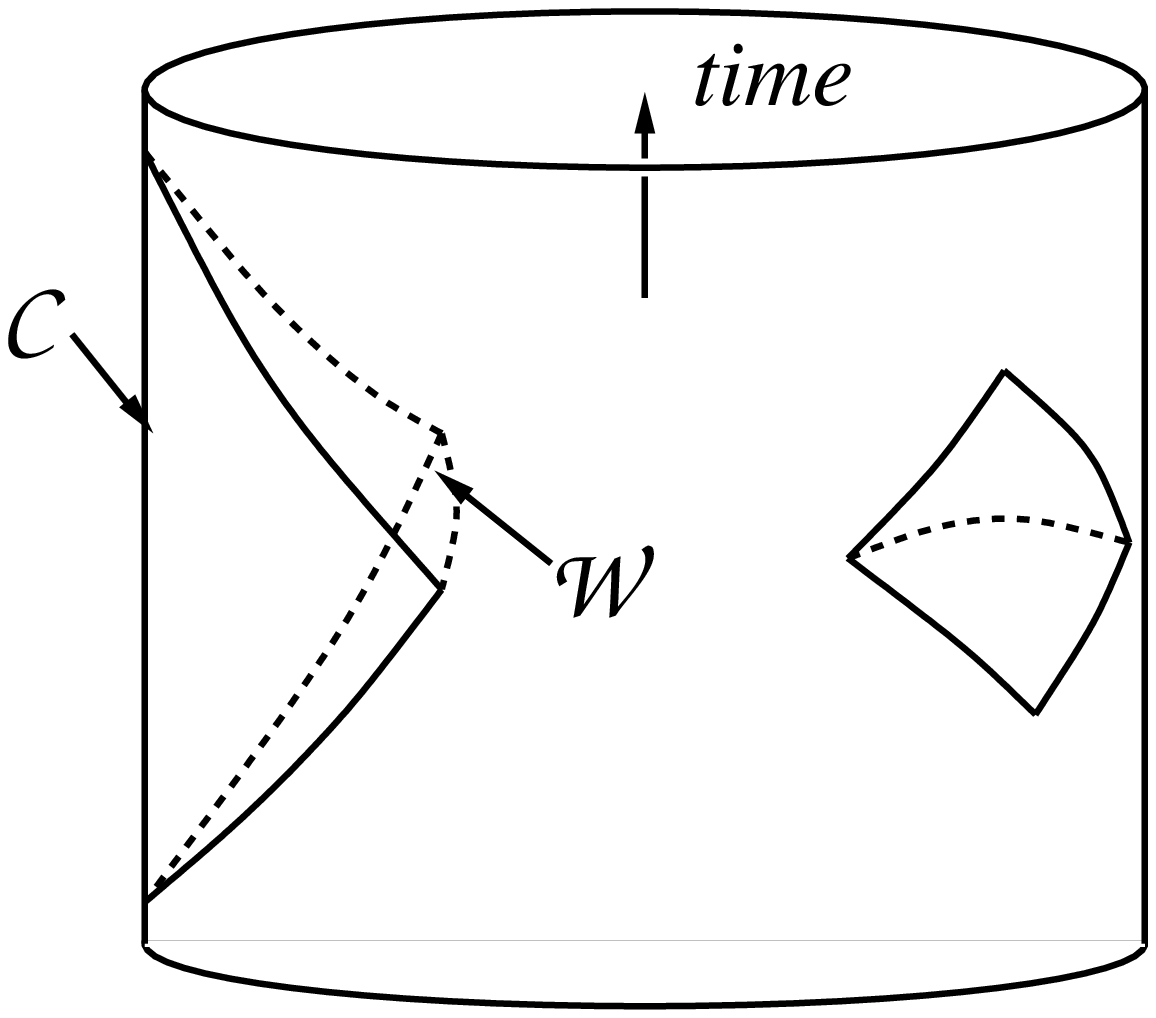}
\caption{\small Wedges in AdS and diamonds on its Minkowskian boundary}
\end{figure}

\nind It is crucial that with this 
choice there exists a bijection 
$\, \gamma : \{ \CW \} \mapsto \{\CC \}$ between these regions which is 
\begin{itemize} 
\item[$\triangleright$] causal: $\gamma(\CW^{\, \prime}) = 
\gamma(\CW)^{\,    \prime}$, where the prime indicates causal
complementation  
\item[$\triangleright$] symmetric: $\gamma(g\CW) = g \,\gamma(\CW)$
  for $\, g\in\mbox{iso}\, \mbox{AdS} = 
  \mbox{conf}\, \partial \mbox{AdS}$, where iso and conf indicate the
  isometry and conformal group of the respective spaces  
\item[$\triangleright$] order preserving: $\gamma(\CW_1) \subset
  \gamma(\CW_2)$ if $\CW_1 \subset \CW_2$.
\end{itemize}

After these geometrical preparations it is straightforward to
establish the desired correspondence between nets on AdS and
$\partial$AdS as well as the corresponding unitary representations of the 
respective symmetry groups and the vacuum states, setting 
\begin{itemize}
\item[$\triangleright$] $\CA_{\, \partial \mbox{\scriptsize AdS}}(\CC) \equiv  
\CA_{\, \mbox{\scriptsize AdS}}(\CW) $ \ for \ $\CW = \gamma^{-1}(\CC)$
\item[$\triangleright$] $U_{\, \partial \mbox{\scriptsize AdS}}(g) \equiv 
U_{\mbox{\scriptsize AdS}}(g)$ 
\item[$\triangleright$] $\Omega_{\, \partial \mbox{\scriptsize AdS}} \equiv 
\Omega_{\mbox{\scriptsize AdS}}$. 
\end{itemize}
Starting from a local, covariant and stable net on AdS one obtains in
this way a local, conformally covariant and stable net on Minkowski
space and {\it vice versa\/} (by reading the defining relations from 
right to left). So one arrives at \cite{Re}
 
\begin{proposition}
There is a one--to--one correspondence between local, covariant
and stable QFT's on AdS and on $\partial$AdS = conformal Minkowski space.
\end{proposition}
It seems plausible that the general idea of holography can also 
be applied to other space--time manifolds with suitable boundaries. \\[3mm]
\nind Transplantation: \\ 
The possibility of identifying quantum field theories on space--times of 
equal dimension (transplantation) has been exemplified in
\cite{BuMuSu} for a special class of Robertson--Walker space--times (RW).
In this approach one makes use of the fact that these space--times 
can be conformally embedded into de Sitter space (dS). Given 
this embedding, one chooses a certain specific family 
$\{ \CC_{\mbox{\scriptsize dS}} \}$ 
of diamond shaped regions in dS and, taking the 
intersection with RW of those regions $\CC_{\mbox{\scriptsize dS}}$
whose edge lies completely in RW, one obtains a corresponding family 
of regions $\{ \CC_{\mbox{\scriptsize RW}} \}$ in RW,
$$\{ \CC_{\mbox{\scriptsize RW}} \equiv \CC _{\mbox{\scriptsize dS}} \cap
\mbox{RW} :\, \mbox{edge}\ \CC _{\mbox{\scriptsize dS}} \subset
{\mbox{RW}} \}.
$$
This construction is indicated in 
Figure 2. We mention as an aside that the
respective regions can also be characterized in an intrinsic
coordinate independent manner. 

\vspace*{-10mm}
\begin{figure}[h]
\epsfxsize130mm
\epsfbox{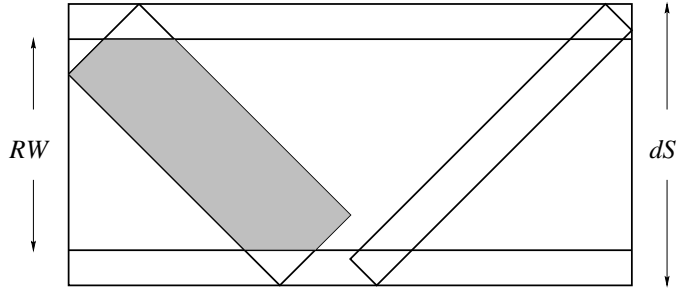}
\caption{\small Penrose diagram with diamonds in RW (left) and dS (right)} 
\end{figure}

\nind Given these regions, there exists a   
unique bijection $\gamma$ : $\{ \CC_{\mbox{\scriptsize dS}} \} \rightarrow
\{ \CC_{\mbox{\scriptsize RW}} \}$ with the property of being  
\begin{itemize}
\item[$\triangleright$] causal: \ 
$\gamma({\CC_{\mbox{\scriptsize dS}}}^{\, \prime}) = 
\gamma(\CC_{\mbox{\scriptsize dS}})_{ }^{\, \prime}$ 
\item[$\triangleright$] symmetric: \
$\dot{g} \, \gamma(\CC_{\mbox{\scriptsize dS}}) 
= \gamma(g \, \CC_{\mbox{\scriptsize dS}})$, \
$g \in \mbox{iso} \, \mbox{dS}.$ 
\end{itemize}
Here $\dot{g}$ denotes the induced action of the elements 
$g$ of the isometry group of de Sitter space, iso dS, 
on the family of regions 
$\{ \CC_{\mbox{\scriptsize RW}} \}$. This action can in general not
be described by a point transformation on RW. But if 
$g \in \mbox{iso} \, \mbox{RW} \subset \mbox{iso} \, \mbox{dS}$
one has $\dot{g} = g$.

\noindent 
On the basis of these geometrical facts it is straightforward to establish a 
one--to--one correspondence between theories on dS and RW, setting 
\begin{itemize}
\item[$\triangleright$] 
$\CA_{\mbox{\scriptsize RW}}(\CC_{\mbox{\scriptsize RW}}) 
\equiv  \CA_{{\mbox{\scriptsize dS}}}(\CC_{\mbox{\scriptsize dS}}) $ \ iff \
$\CC_{\mbox{\scriptsize RW}} = \gamma(\CC_{\mbox{\scriptsize dS}})$ 
\item[$\triangleright$]
$U_{\mbox{\scriptsize RW}}(\dot{g}) \equiv U_{\mbox{\scriptsize
      dS}}(g)$,  \ $g \in \mbox{iso} \, \mbox{dS}$
\end{itemize}
An immediate consequence of the geometrical properties of $\gamma$ is 
\cite{BuMuSu}
\begin{proposition} 
Each local, covariant {\it dS}--theory can be mapped to 
a local, covariant (ultra symmetric) {\it RW}--theory and 
{\it vice versa}.
\end{proposition}
Here the term ``ultra symmetric'' means that the resulting RW theory 
exhibits, besides the familiar space--time isometries,  
an aditional geometrical symmetry which is {\it not\/} induced by point
transformations. This situation differs from the AdS/CFT correspondence, 
where the Minkowskian theory on the boundary of AdS is conformally invariant. 
 

\section{Modular construction of local nets} 
An intriguing recent result in algebraic quantum
field theory is the insight that local nets can be constructed
from a few local algebras in suitable ``relative positions'' 
\cite{Wi0,Wi1,Wi2,KaWi}.
It is of interest in this context that the local von Neumann algebras are 
universal (model independent) objects: they are generically  
\mbox{isomorphic} to the {\it unique} hyperfinite type III$_1$ factor 
\cite{BuDaFr}.

So the starting point of this novel construction is a concrete and
well--studied algebra, denoted by $\CM$ in the 
following; the second ingredient is a standard 
(cyclic and separating) vector $\Omega$ for $\CM$
in the underlying Hilbert space. Given these quantities, one 
can consistently define a conjugation
$$ S_{\CM} : M \, \Omega \mapsto M^* \, \Omega.$$
It is an anti--linear closable operator whose closure is denoted  
by the same symbol and whose polar decomposition has the form 
$$ S_{\CM} = J_{\CM}^{ } \, \Delta_{\CM}^{1/2}.$$
Here $J_{\CM}\,$ is an anti--unitary operator, 
called {\it modular conjugation},
and the positive selfadjoint operator $\Delta_{\CM}$ is 
the {\it modular operator}. The corresponding unitary group 
$\{ {\Delta_{\CM}}^{\! -is} \}_{s \in \RR}$ 
is called {\it modular group}. Irrespective of the choice of 
$\CM$ and $\Omega$ within the above limitations, 
there hold the following basic relations 
established by Tomita and Takesaki:
\begin{itemize}
\item[$\triangleright$]  
${\Delta_{\CM}}^{\! -is} \, \CM \, {\Delta_{\CM}}^{is} =
\CM, \quad s \in \RR$ 
\item[$\triangleright$]  
$J_{\CM}^{ } \, \CM \, {J_{\CM}}^{-1} = \CM^{\, \prime}.$
\end{itemize}
Based on these facts, Wiesbrock introduced in \cite{Wi1} the notion of 
{\it half--sided modular inclusion\/} of a von Neumann
algebra $\CN \subset \CM$, where $\CN$ likewise 
has $\Omega$ as standard vector, by posing the condition  
$$
{\Delta_{\CM}}^{\! -is} \, \CN \, {\Delta_{\CM}}^{\, is} 
\subset \CN, \quad s \in \RR_+. \eqno(*) 
$$
The unitary group $U$, obtained by the Trotter product formula 
$$ U(t) \equiv \lim_{n \rightarrow \infty} \,
({\Delta_{\CM}}^{\! -it/2\pi n} \, {\Delta_{\CN}}^{\, it/2 \pi n})^n,
\quad t \in \RR,$$
then has the following properties \cite{Wi0}. 
\begin{proposition} Let $\CN \subset \CM$ be a half--sided modular inclusion. 
The corresponding modular groups generate a unitary 
representation of \, $\RR_+\ltimes \RR$ such that
\begin{itemize}
\item[a)]  ${\Delta_{\CM}}^{\! -is} \, U(t) =
U(e^{2 \pi s} \, t) \, {\Delta_{\CM}}^{\! -is}, \quad s,t \in \RR$
\item[b)]  $J_{\CM} \, U(t) \, {J_{\CM}}^{-1} = U(-t)$
\item[c)] the spectrum of the generator of $U$ is contained in 
$\RR_+$  
\item[d)] $\CN = U(1) \, \CM \, U(1)^{-1}$.
\end{itemize}
\end{proposition}
An analogous result holds if $\RR_+$ is replaced by $\RR_-$ in $(*)$.
Another useful concept, characterizing the relative position of von Neumann
algebras, is the notion of {\it modular intersection} \cite{Wi2}: two von
Neumann algebras $\CM$, $\CN$ are said to have modular intersection
if $\CM \cap \CN$ is half--sided modular in $\CM$ and $\CN$,
respectively. Similarly to the situation discussed in the preceding 
proposition, the modular groups can be shown to generate a unitary 
representations of a Lie group in the latter case as well. 
These general mathematical facts have immediate applications in
algebraic quantum field theory.

\vspace*{2mm}
\nind 1. Local nets on $\RR$ \ \cite{Wi1} \\[2mm]
Any modular inclusion fixes a local, covariant and stable 
net of local algebras on the light ray $\RR$.
One first assigns algebras to half lines, setting 
$$\CA(\RR_+ + x) \equiv U(x) \CM U(x)^{-1}, \quad
\CA(\RR_- + y) \equiv U(y) {\CM}^{\, \prime} U(y)^{-1}.$$
The algebras corresponding to arbitrary intervals $I = [x, y] \subset \RR$ are
given by 
$$ \CA (I) \equiv \CA(\RR_+ + x)  \, \cap \, \CA(\RR_- + y).$$
It then follows from the preceding proposition and the basic relations
of Tomita--Takesaki--Theory that the 
net $I \mapsto \CA(I)$ on $\RR$ is local (operators localized in
disjoint intervals commute), $\RR_+ \ltimes \RR$--covariant and 
stable ($\Omega$ being a ground state for $U$). Thus, given two algebras 
in suitable relative position, one can construct 
a full chiral quantum field theory. (In a similar way one sees that
any modular intersection fixes a conformally invariant quantum field
theory on the compactified light ray S$^1$.)

\vspace*{2mm}
\nind 2. Local nets on $\RR^2$ \ \cite{Wi1} \\[2mm]
In order to obtain in a similar manner 
local nets on higher dimensional space--times, one has to
proceed from a larger set of algebras in appropriate relative 
positions. In the case of two--dimensional Minkowski space,  
one starts from three algebras, forming two half--sided 
modular inclusions $\CN_\pm \subset \CM$. By the preceding 
proposition one then has two unitary groups 
$U_\pm (x_\pm)$, where $x_\pm$ are interpreted as light cone
coordinates of $x\in\RR^2$. These groups are assumed to 
commute, 
$$ U(x) \equiv U_+(x_+) \, U_-(x_-) \stackrel{!}{=}
 U_-(x_-) \, U_+(x_+).$$
In this way one obtains a unitary positive energy representation $U$ of the 
translations $\RR^2$ on which the modular group of $\CM$ acts like
a Lorentz transformation,  
$$ {\Delta_{\CM}}^{\! -is} U(x) {\Delta_{\CM}}^{is} = 
U_+(e^{2 \pi s} \, x_+) U_-(e^{- 2 \pi s} \, x_-) = U(\Lambda(s) x). 
$$
One then can proceed as in the preceding case and define algebras for 
wedge shaped regions of the form $\CW = \{x \in \RR^2: x_1 > |x_0|\}$, setting
$$
\CA (\CW + x) \equiv  U(x) \, {\CM} \, U(x)^{-1}, \quad
\CA ({\CW}^{\, \prime} + y) \equiv  U(y) \, {\CM}^{\,\prime} \,
U(y)^{-1}.
$$
The algebras associated with diamonds 
$\CC_{x,y} = (\CW +x) \cap (\CW^\prime + y)$, cf. Figure 3, 
are obtained by setting 
$$ \CA (\CC_{x,y}) \equiv  \CA (\CW + x) \cap \CA ({\CW}^{\, \prime} + y).$$ 
In this way one arrives at a local, Poincar\'e--covariant net on 
$\RR^2$ where $\Omega$ describes the vacuum vector.

\begin{figure}[h]
\epsfxsize50mm
\hspace*{55mm}
\epsfbox{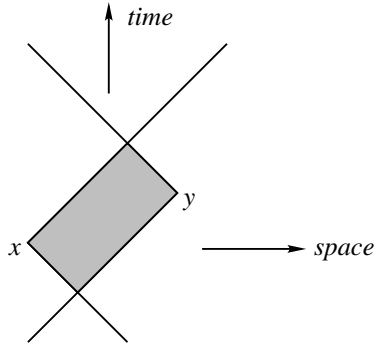}
\caption{\small A diamond obtained as intersection of wedges}
\end{figure}

\vspace*{2mm}
\nind 3. Local nets on $\RR^d$, $d=3,4$ \ \cite{Wi2,KaWi} \\[2mm]
The construction of local nets from a few algebras was recently 
extended to three and four--dimensional Minkowski space in
\cite{Wi2,KaWi}. The crucial and difficult 
step in this approach is the formulation
of conditions which guarantee that the modular groups affiliated with
the algebras and the underlying vector $\Omega$ generate 
representations of the Poincar\'e group $\CP_+^\uparrow$. 
\begin{proposition}  
Any family of algebras $\CM_1, \dots \CM_{d(d-1)/2 +1 }$ 
in suitable modular positions fixes a local, $\CP_+^\uparrow$--covariant 
net on $\RR^d$ with vacuum state $\Omega$.
\end{proposition}
We refrain from giving here the precise conditions on the 
algebras and only note that, in analogy to the cases discussed before, 
the corresponding modular groups generate 
representation of $\CP_+^\uparrow$ with positive energy. 
The algebras $\CM_{i}$ can consistently be assigned to wedge 
regions in $\RR^d$ and the local algebras associated with diamonds are
defined by taking intersections. The converse of this statement is a
well--known theorem by Bisognano and Wichmann \cite{BiWi}, cf.\ also 
\cite{Bo}.

These intriguing results should admit a generalization to other
space--time manifolds, cf.\ also \cite{BuDrFlSu} for a related approach.
Moreover, they seem to be of relevance for the classification of local
nets and are possibly a step towards a novel, completely algebraic
approach to the construction of local nets. 


\section{Conclusion}  
The preceding account of recent results in AQFT illustrates the 
role of this approach in relativistic quantum field theory: it is a
concise framework which is suitable for the development of new constructive 
schemes, the mathematical implementation of physical concepts and ideas,
the elaboration of general computational methods and the clarification 
of the relation between different theories as well as 
their structural analysis and classification. 
So this framework complements the 
more concrete approaches to relativistic quantum field theory and
thereby contributes to the understanding and mathematical consolidation 
of this important area of mathematical physics.

\end{document}